\documentclass[12pt]{iopart}

\expandafter\let\csname equation*\endcsname\relax    %Solve conflict iopart/amsmath
\expandafter\let\csname endequation*\endcsname\relax %Solve conflict iopart/amsmath

\usepackage[colorlinks=true,urlcolor=blue,citecolor=blue,linkcolor=red]{hyperref}

\usepackage{amsmath,amsfonts,amssymb,bm}
\usepackage{bbold} %for the identity matrix

\usepackage{graphicx}

\usepackage{cite}

\begin{document}

\title[Dynamics of interacting multispecies quantum gases]{Efficient numerical description of the dynamics of interacting multispecies quantum gases}

\author{Annie Pichery$^{1,2}$, Matthias Meister$^{3}$, Baptist Piest$^{1}$, Jonas Böhm$^{1}$, Ernst Maria Rasel$^{1}$, Eric Charron$^{2}$, Naceur Gaaloul$^{1}$}

\address{$^{1}\,$Leibniz University of Hanover, Institute of Quantum Optics, QUEST-Leibniz Research School, Hanover, Germany.}
\address{$^{2}\,$Université Paris-Saclay, CNRS, Institut des Sciences Moléculaires d’Orsay, F-91405 Orsay, France.}
\address{$^{3}\,$German Aerospace Center (DLR), Institute of
Quantum Technologies, Ulm, Germany.}

\vspace{10pt}
\begin{indented}
\item[]\today
\end{indented}

\begin{abstract}
We present a highly efficient method for the numerical solution of coupled Gross-Pitaevskii equations describing the evolution dynamics of a multi-species mixture of Bose-Einstein condensates in time-dependent potentials. This method, based on a grid-scaling technique, compares favorably to a more standard but much more computationally expensive solution based on a frozen-resolution grid. It allows an accurate description of the long-time behavior of interacting, multi-species quantum mixtures including the challenging problem of long free expansions relevant for microgravity and space experiments. We demonstrate a successful comparison to experimental measurements of a binary Rb-K mixture recently performed with the payload of a sounding rocket experiment.
\end{abstract}

\section{Introduction}

Degenerate atomic mixtures are a very rich system and have inspired a wealth of theoretical~\cite{PuPRL1998, TrippenbachJPhysB2000, RiboliPRA2002, NavarroPRA2009, BalazPRA2012, VidanovicNJP2013, PattinsonPRA2013, PoloPRA2015, LeePRA2016, LeeNJP2018, CorgierNJP2020, WolfPRA2022, MeisterQST2023} and experimental research \cite{FerrariPRL2002, ModugnoPRL2002, FerlainoPRA2006, FerlainoPRA2006Erratum, PappPRL2008, RonenPRA2008, ThalhammerPRL2008, BurchiantiPRA2018, WilsonPRA2021, WilsonPRRes2021, JiaPRL2022, CavicchioliPRRes2022}. They may consist of two (or more) components, which can be the internal spin states of a single species of a Bose-Einstein condensate \cite{BalazPRA2012}, two isotopes of a single species \cite{RonenPRA2008}, or two different species \cite{PoloPRA2015, PattinsonPRA2013, WilsonPRA2021, CavicchioliPRRes2022}. In recent years, interest in binary mixtures has spread from pure quantum gas physics to metrology, and in particular to their use in high-precision atomic interferometry experiments. Indeed, two atomic species could be used as input states of a dual-atom interferometer to test fundamental principles such as the universality of free fall. In this context, recent proposals \cite{AguileraClassQuGrav2014, HenselEuPhysJD2021, BattelierExpAstro2021, AhlersArXiv2022stequest} predict the manipulation of quantum mixtures over large distances, in weak traps or in free fall, which could last tens of seconds, thus increasing the sensitivity of the atomic sensor \cite{Berman1997}. These time scales challenge the current state of the art in computational resources, since one has to solve at least a coupled set of Gross-Pitaevskii equations in the mean-field framework to reproduce the complex dynamics driven by the interaction of the two quantum gases. Indeed, approaches based on a Thomas-Fermi approximation or dimension reduction, e.g. by adopting spherical coordinates, remain specific to a few examples of experimental settings and cannot be generalized to time-dependent situations where the interactions lead to exotic states or symmetry breaking. 

In this work, we generalize grid scaling techniques developed in the single-species case \cite{CastinDumPRL1996, Shlyapnikov1997, EckartThesis2008, MeisterAAMOP2017, Bradley2022} to the multi-species case in order to efficiently handle the transport, or expansion dynamics of these systems. This method turns out to be numerically very efficient and allows access to time regimes that are inaccessible with static grid arrangements. We expect this scheme to be instrumental in describing quantum gases at long expansion times as proposed in microgravity or space experiments \cite{AvelineNature2020, Gaaloul2022, FryeEPJ2021, MuntingaPRL2013, Deppner2021, Becker2018, Barrett2016, condon2019, Barrett2022, Lotz2023}. We illustrate our findings by solving the ground states and dynamics of mixtures of K-41 and Rb-87, as these are the systems considered in these projects. Finally, to validate the theoretical treatment, we compare our results with the detected images of BEC mixtures recorded by the MAIUS-2 sounding rocket team during the ground tests of its payload \cite{PiestThesis2021}. We find an excellent agreement and prove the relevance of the developed toolbox for microgravity and space investigations.

This paper is organized as follows: Section\;\ref{sec:theory} is devoted to the development of our theoretical approach aimed at solving the coupled multi-species BEC dynamics in a general 3D time-dependent trap or during a free expansion stage. In section\;\ref{sec:applications}, we first present two generic examples: the transport of a Rb-K two-species condensate in microgravity and its free expansion in the presence of gravity. The results obtained with our grid-scaling approach are systematically compared with the more standard, but much more time-consuming, calculations obtained with a fixed grid. In the same section, we also compare the predictions of our efficient numerical approach with experimental test measurements performed on the ground with the MAIUS-2 sounding rocket platform. Finally, a summary and conclusion are given in section\;\ref{sec:conclusion}.

\section{Theoretical Approach}
\label{sec:theory}

\subsection{Theoretical Model}

At zero temperature and within the mean-field approximation, the time evolution of a multispecies mixture of Bose-Einstein condensates is described by the time-dependent coupled Gross-Pitaevskii equations
\begin{equation}
\label{eq_TDCGPE}
i\hbar\,\partial_t \Psi_j(\mathbf{r},t) = \left[-\dfrac{\hbar^2}{2 m_j}\bm{\nabla}^2_{\mathbf{r}} + U_j(\mathbf{r},t) + \sum_{j'=1}^{n_{sp}}N_{j'}\,g_{jj'}|\,\Psi_{j'}(\mathbf{r},t)|^2 \right] \Psi_j(\mathbf{r},t)
\end{equation}
where $j$ and $j' = 1, 2, ..., n_{sp}$ are the labels associated with the $n_{sp}$ different atomic species. In this expression $\mathbf{r} = (x, y, z)^{T}$ denotes the position vector in a fixed frame of reference, and $^T$ is a simple notation used here to indicate transposition. $\Psi_j(\mathbf{r},t)$ is the normalized wave function of the species number $j$, of mass $m_j$. $N_j$ and $U_j(\mathbf{r},t)$ are the atom number and the external potential of species $j$. The scattering amplitudes $g_{jj'}$ are related to the corresponding $s$-wave scattering lengths $a_{jj'}$ by the relation
\begin{equation}
g_{jj'} = \frac{2\pi\hbar^2 a_{jj'}}{m_{jj'}}\,,    
\end{equation}
where $m_{jj'}$ denotes the reduced mass
\begin{equation}
m_{jj'} = \frac{m_{j}m_{j'}}{m_{j}+m_{j'}}\,.
\end{equation}

\begin{figure}[!t]
\centering
\includegraphics[width=0.5\textwidth]{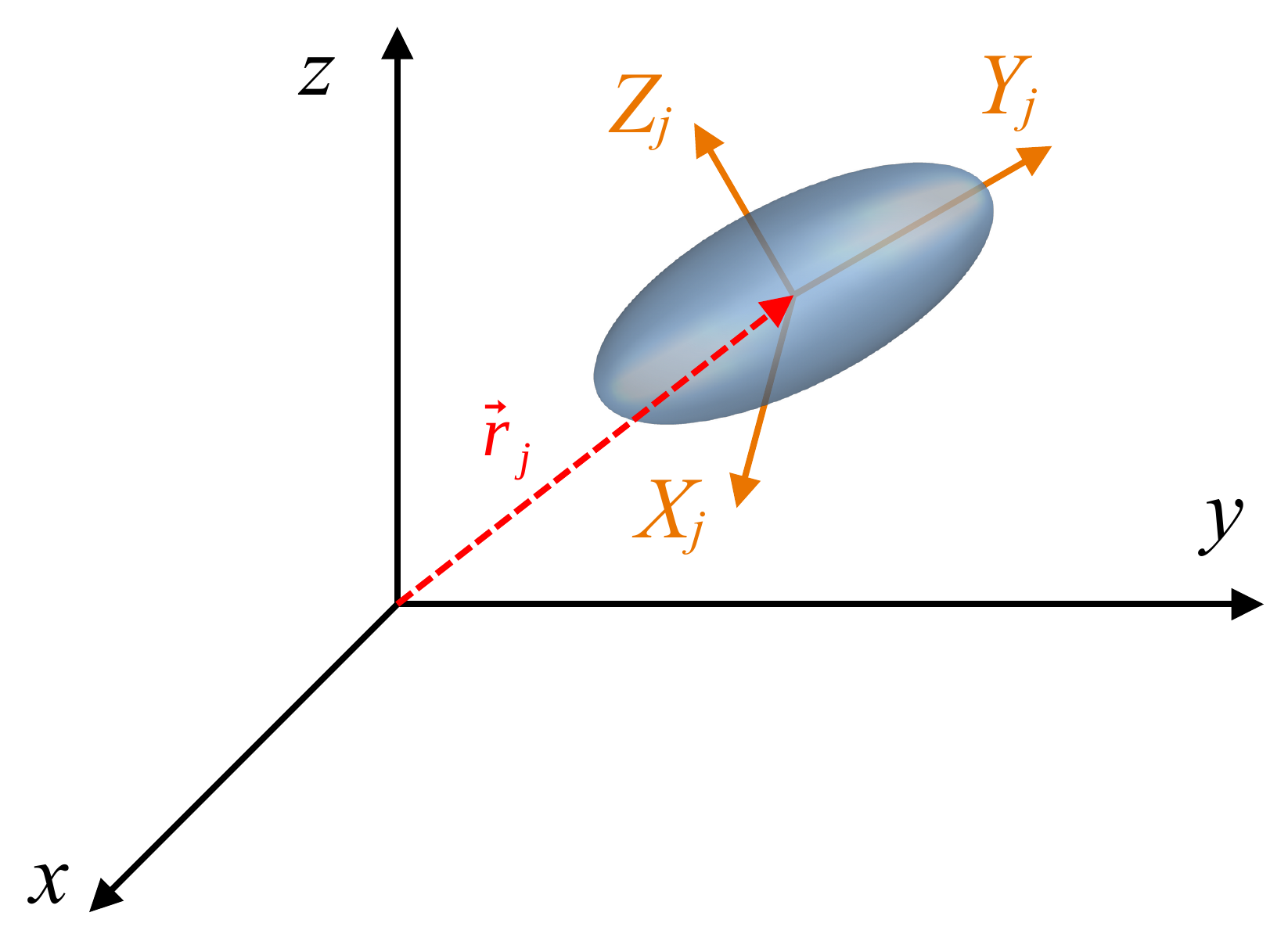}
\caption{Schematic representation of the condensate (in blue) associated with the species $j$, centered on the point of coordinates $\mathbf{r}_j=(x_j,y_j,z_j)^T$ in the fixed reference frame $(x,y,z)$.  The reference frame $(X_j,Y_j,Z_j)$ associated with the eigenaxes of the harmonic trap $V_j$ is shown in orange.}
\label{Fig:Rotated_Frame}
\end{figure}

In the following we will consider that the multispecies condensate is trapped in a general external potential given by the functions $U_j(\mathbf{r},t)$, that we decompose into the sum of a harmonic and an anharmonic part, according to
\begin{equation}
U_j(\mathbf{r},t) = V_{j}(\mathbf{r},t) + W_{\!j}(\mathbf{r},t)\,,
\end{equation}
where
\begin{equation}
V_j(\mathbf{r},t) = \frac{1}{2}\,m_{j}\,\big(\mathbf{r}-\mathbf{r}_j(t)\big)^T\,\bm{\Omega}_j^2(t)\,\big(\mathbf{r}-\mathbf{r}_j(t)\big)\,.
\label{Eq:Vharmj}
\end{equation}
In this expression, $\mathbf{r}_j(t)=(x_j(t),y_j(t),z_j(t))^T$ is the position of the trap minimum for species $j$ at time $t$ in the fixed reference frame. The axis and coordinate systems introduced here are shown schematically in Fig.\,\ref{Fig:Rotated_Frame}. We assume that at each time $t$ the harmonic traps $V_j(\mathbf{r},t)$ associated with the different species are characterized by eigenaxes pointing in the directions $X_j(t)$, $Y_j(t)$ and $Z_j(t)$. The unitary rotation matrix that allows to pass from the particular system of eigenaxes $(X_j(t), Y_j(t), Z_j(t))$ to the fixed frame of reference $(x, y, z)$ at time $t$ is denoted by $\mathbf{M}_j(t)$. The $3 \times 3$ squared harmonic frequency matrices $\bm{\Omega}_j^2(t)$ are then defined in the fixed reference frame $(x,y,z)$ as
\begin{equation}
\bm{\Omega}_j^2(t) = \mathbf{M}_j(t)
\begin{pmatrix}
\omega_{j,X_j}^2(t) & 0 & 0 \\ 
0 & \omega_{j,Y_j}^2(t) & 0 \\ 
0 & 0 & \omega_{j,Z_j}^2(t) 
\end{pmatrix}
\mathbf{M}_j(t)^T\,.
\label{Eq:Omega_j}
\end{equation}
The eigenvalues of $\bm{\Omega}_j^2(t)$ thus coincide with the squared instantaneous eigenfrequencies $\omega_{j,X_j}^2(t)$, $\omega_{j,Y_j}^2(t)$ and $\omega_{j,Z_j}^2(t)$ of the traps along their principal axes $(X_j(t), Y_j(t), Z_j(t))$. In our study, the calculation of the dynamics is carried out in the reference frame corresponding to the eigenaxes of a particular species, the species of index $j=j^*$, which in principle can be chosen freely. In all that follows, we will assume that for this particular species, the eigenaxes of the trap do not rotate during the dynamics. Thus, the rotation matrix $\mathbf{M}_{j^*}(t)$ will simply be denoted $\mathbf{M}_{j^*}$, and will be assumed to be independent of time. In practice, this approach can be used as long as the rotation of these eigenaxes is sufficiently slow so that the effect of non-inertial forces due to the rotation of the eigenaxes associated with this reference species $j^*$ can be neglected. This is the case in most situations, especially when the rotation is slow enough not to induce the appearance of vortices, as shown for example in Refs.\,\cite{HechenblaiknerPRL2002, EdwardsPRL2002, MeisterAAMOP2017}.

In the reference frame associated with the eigenaxes of the species $j^*$, the matrix of the squared harmonic frequencies associated with each species $j$ writes
\begin{equation}
\bm{\Omega}_j^{\prime\,2}(t) \,=\, \mathbf{M}_{j^*}^T\;
\bm{\Omega}_j^2(t)\;
\mathbf{M}_{j^*}\,.
\label{Eq:Omega_jprime}
\end{equation}
This matrix is generally a non-diagonal but symmetric matrix. In fact, the matrix $\bm{\Omega}_j^{\prime\,2}(t)$ is diagonal only if the trap associated with the species $j$ has the same principal axes as the trap associated with the reference species $j^*$. We can also verify by a simple use of equations (\ref{Eq:Omega_j}) and (\ref{Eq:Omega_jprime}) that $\bm{\Omega}_{j^*}^{\prime\, 2}(t)$ is the diagonal matrix of the squares of the instantaneous eigenfrequencies $\omega_{j^*,X_{j^*}}^2\!(t)$, $\omega_{j^*,Y_{j^*}}^2\!(t)$ and $\omega_{j^*,Z_{j^*}}^2\!(t)$, as expected.

\subsection{Moving the Grid}
\label{sec:theory:com_scaling}

When a temporal variation in the position and/or frequency of the traps induces a displacement of the multispecies condensate, and when the amplitude of this displacement is large, it can be extremely useful to shift the grid (or equivalently the reference frame) in which the dynamics is computed in order to save computational time. This is especially important when implementing condensate transport over distances significantly larger than the characteristic size of the condensate, as in the recent space atom chip manipulation of a BEC of Ref. \cite{Gaaloul2022}. For this purpose, we impose on the computational grid to follow the global displacement dictated by the classical evolution of the condensate center of mass of the reference species $j=j^*$. This approach consists in a further development of the treatments that have already been presented in the Refs.\,\cite{EckartThesis2008, MeisterAAMOP2017, MeisterThesis2019, MeisterQST2023}. The change of variable associated with this transformation results in the introduction of the new coordinate
\begin{equation}
\mathbf{R} = \mathbf{r} - \mathbf{r}_{\mathrm{cm},j^*}(t)
\end{equation}
where $\mathbf{r}_{\mathrm{cm},j^*}(t)$ denotes the classical position of the center of mass of the condensate associated with the species $j^*$ at time $t$, computed by simply solving Newton's equation for a classical particle of mass $m_{j^*}$ initially at rest and subjected to the time-dependent harmonic potential $V_{j^*}(\mathbf{r},t)$ of Eq.\,(\ref{Eq:Vharmj}).This allows us to define the quantum displacement operator in coordinate and momentum space
\begin{equation}
\hat{D}_{j}(t) = \exp\left(i\,\big[\,\mathbf{k}_{\mathrm{cm},j}(t)\cdot\hat{\mathbf{r}} - \mathbf{r}_{\mathrm{cm},j^*}(t)\cdot\hat{\mathbf{k}}\,\big]\right)
\end{equation}
where $\hat{\mathbf{k}}=-i\,\bm{\nabla}_{\!\mathbf{r}}$ and $\mathbf{k}_{\mathrm{cm},j}(t) = \mathbf{p}_{\mathrm{cm},j}(t)/\hbar$. In this expression, we find the classical momentum $\mathbf{p}_{\mathrm{cm},j}(t) = m_{j}\, \dot{\mathbf{r}}_{\mathrm{cm},j}(t)$ with $\dot{\mathbf{p}}_{\mathrm{cm},j}(t) = - m_{j}\, \bm{\Omega}_{j}^{\prime\,2}(t)\,[\mathbf{r}_{\mathrm{cm},j}(t)-\mathbf{r}_{j}(t)]$. Following~\cite{MeisterAAMOP2017}, the unitary transformation
\begin{equation}
\Psi_{j}(\mathbf{r},t) = e^{i\,S_{j}(t)/\hbar}\,\hat{D}_{j}(t)\,\Psi_{j}^D(\mathbf{R},t)
\end{equation}
with an adapted global phase $S_j(t)$ which satisfies
\begin{equation}
\frac{dS_{j}}{dt} =
- \dot{\mathbf{p}}_{\mathrm{cm},j} \cdot \mathbf{r}_{\mathrm{cm},j^*}
+ \frac{1}{2} \frac{d}{dt} \big[ \mathbf{r}_{\mathrm{cm},j^*} \cdot \mathbf{p}_{\mathrm{cm},j} \big]
- \frac{\mathbf{p}_{\mathrm{cm},j}^2}{2m_j}
- \dfrac{m_{j^*}}{2}(\mathbf{r}_{\mathrm{cm},j^*} - \mathbf{r}_{j})^T\bm{\Omega}_{j}^{\prime\,2}(\mathbf{r}_{\mathrm{cm},j^*}-\mathbf{r}_{j})
\end{equation}
leads to the following transformed Gross-Pitaevskii equation for the species $j$, written in the frame associated with the motion of the classical center of mass of the species $j^*$
\begin{multline}
i\hbar\,\partial_t\Psi_{j}^D(\mathbf{R},t) = \bigg[
-\dfrac{\hbar^2}{2m_j}\bm{\nabla}_{\!R}^2
+ \frac{m_j}{2}\,\mathbf{R}^T\,\bm{\Omega}_{j}^{\prime\,2}(t)\,\mathbf{R}
+ \overline{W}_{\!j}(\mathbf{R},t)+ V_j^{\mathrm{cor}}(\mathbf{R},t) \\
+\sum_{j'=1}^{n_{sp}}N_{j'}\,g_{jj'}|\,\Psi_{j'}^D(\mathbf{R},t)|^2
\bigg] \Psi_{j}^D(\mathbf{R},t)\,,
\label{Eq:GPE2_COM}
\end{multline}
where $\overline{W}_{\!j}(\mathbf{R},t) = W_{\!j}(\mathbf{r}-\mathbf{r}_{\mathrm{cm},j^*},t)$ and where $V_j^{\mathrm{cor}}(\mathbf{R},t)$ is a linear correction term written as
\begin{equation}
V_j^{\mathrm{cor}}(\mathbf{R},t) = 
  m_j\,\Big[(\mathbf{r}_{\mathrm{cm},j^*}-\mathbf{r}_{j})^T\,\bm{\Omega}_{j}^{\prime\,2}(t)
- (\mathbf{r}_{\mathrm{cm},j^*}-\mathbf{r}_{j^*})^T\,\bm{\Omega}_{j^*}^{\prime\,2}(t)\Big]\,\mathbf{R}\,.
\label{Eq:Vcor}
\end{equation}
Equation (\ref{Eq:GPE2_COM}) shows as a unique coordinate the translated coordinate $\mathbf{R}=\mathbf{r}-\mathbf{r}_{\mathrm{cm},j^*}(t)$, enabling us to see that the new computational grid follows the global motion of the center of mass of the condensate associated with the reference species  $j^*$. In practice, if we now substitute $j$ for $j^*$ in Eq.\,(\ref{Eq:Vcor}), we see that the correction term (\ref{Eq:Vcor}) disappears and as a consequence Eq.\,(\ref{Eq:GPE2_COM}) reduces to
\begin{multline}
i\hbar\,\partial_t\Psi_{j^*}^D(\mathbf{R},t) = \bigg[ -\dfrac{\hbar^2}{2m_{j^*}}\bm{\nabla}_{\!R}^2 + \frac{m_{j^*}}{2} \mathbf{R}^T\bm{\Omega}_{j^*}^{\prime\,2}(t)\mathbf{R} + \overline{W}_{\!j^*}(\mathbf{R},t) \\
+\sum_{j'=1}^{n_{sp}}N_{j'}\,g_{j^*j'}|\,\Psi_{j'}^D(\mathbf{R},t)|^2 \bigg] \Psi_{j^*}^D(\mathbf{R},t) \,.
\label{Eq:GPE1_COM}
\end{multline}
Eq.\,(\ref{Eq:GPE2_COM}) can thus be considered as a general equation applicable to any species, whether or not it is the reference species in the displacement operation being performed.

\subsection{Expanding or Compressing the Grid}
\label{sec:theory:size_scaling}

If the condensate size varies significantly during the dynamics, it may also be useful to compress or expand the grid accordingly during the course of the propagation to save computational time. This approach is especially important when considering a free expansion of the condensate. To define the time-dependent scaling factors applied to the computational grid, we choose the same reference species as before, corresponding to the index $j=j^*$, and we define a new rescaled coordinate $\bm{\xi}$ satisfying
\begin{equation}
\bm{\Lambda}(t)\,\bm{\xi} = \mathbf{R}
\end{equation}
where
\begin{equation}
\bm{\Lambda}(t) = \begin{pmatrix}
\lambda_{X_{j^*}}\!(t) & 0 & 0 \\ 
0 & \lambda_{Y_{j^*}}\!(t) & 0 \\ 
0 & 0 & \lambda_{Z_{j^*}}\!(t)
\end{pmatrix}
\label{Eq:Lambda}
\end{equation}
is a diagonal matrix whose elements are three scalar and adimensional time-dependent scaling functions $\lambda_{X_{j^*}}\!(t)$, $\lambda_{Y_{j^*}}\!(t)$ and $\lambda_{Z_{j^*}}\!(t)$ that we apply to the three coordinates associated with the eigenaxes of the trap experienced by the species number $j^*$. Since the definition of these scaling functions is arbitrary, we chose to force the computational grid to compress or expand according to the dynamics that can be predicted by the so-called ``scaling law'' approximation obtained in the Thomas-Fermi regime\,\cite{CastinDumPRL1996, Shlyapnikov1997}. For a single species BEC with a high number of atoms such that the Thomas-Fermi approximation holds, one can indeed use a classical scaling approximation to describe the 3D size evolution of the BEC in a time-dependent harmonic trap. This amounts to solving the differential equation (written here in a matrix form for a diagonal scaling matrix $\bm{\Lambda}(t)$)
\begin{equation}
\bm{\Lambda}(t)\ddot{\bm{\Lambda}}(t) + \bm{\Omega}_{j^*}^{\prime\,2}(t)\,\bm{\Lambda}^{\!2}(t) = \frac{\bm{\Omega}_{j^*}^{\prime\,2}(0)}{\det[\bm{\Lambda}(t)]}
\label{Eq:ScalingLaw}
\end{equation}
where $\det[\bm{\Lambda}(t)]$ stands for the determinant of the matrix $\bm{\Lambda}(t)$ of Eq.(\ref{Eq:Lambda}). Provided that at time $t=0$ the initial conditions verify $\bm{\Lambda}(0)=\mathbb{1}$ and $\dot{\bm{\Lambda}}(0)=\mathbb{0}$, the scaling factors $\lambda_I(t)$ with $I \in \{X_{j^*},Y_{j^*},Z_{j^*}\}$ usually give a good estimate of the evolution of the BEC size in the three directions $\{X_{j^*},Y_{j^*},Z_{j^*}\}$.

To take into account the introduction of the scaled coordinate $\bm{\xi}$, inspired by\,\cite{MeisterAAMOP2017} we perform the following unitary transformation to the wave function associated with each reference species $j$
\begin{equation}
\Psi_{j}^D(\mathbf{R},t) =
\frac{e^{\frac{i}{\hbar}\big[\bm{\xi}^T\mathbf{A}_j(t)\,\bm{\xi}\,-\,\beta_j(t)\big]}}{\sqrt{\det[\bm{\Lambda}(t)]}}\,\Psi_{j}^S(\bm{\xi},t) \,,
\end{equation}
where
\begin{subequations}
\begin{align}
\mathbf{A}_j(t) & = \frac{1}{2}\,m_j\,\bm{\Lambda}(t)\,\dot{\bm{\Lambda}}(t)\,,\\
\beta_j(t)      & = \int_0^t \frac{\mu_j}{\det[\bm{\Lambda}(t')]}\,dt'\,.
\end{align}
\end{subequations}
This transformation leads to an adapted set of coupled time-dependent Gross-Pitaevskii equations for all species, which reads
\begin{multline}
i\hbar\,\partial_t\Psi_{j}^S(\bm{\xi},t) = \bigg[
- \dfrac{\hbar^2}{2m_j}
\bm{\nabla}_{\!\bm{\xi}}^T\bm{\Lambda}^{\!-2\,}\bm{\nabla}_{\!\bm{\xi}}
+ \dfrac{m_j}{2}\,\bm{\xi}^T\bm{\Lambda}^T \bigg( \bm{\Omega}_j^{\prime\,2}(t) - \bm{\Omega}_{j^*}^{\prime\,2}(t)\bigg)\bm{\Lambda}\,\bm{\xi} + V_j^{\mathrm{cor}}(\bm{\Lambda}\,\bm{\xi},t)\\
+ \overline{W}_{\!j}(\bm{\Lambda}\,\bm{\xi},t)
+ \frac{\frac{m_j}{2}\,\bm{\xi}^T\bm{\Omega}_{j^*}^{\prime\,2}(0)\,\bm{\xi}
+ \sum_{j'}N_{j'}\,g_{jj'}|\,\Psi_{j'}^S(\bm{\xi},t)|^2 - \mu_j}{\det[\bm{\Lambda}(t)]} \bigg]\,\Psi_{j}^S(\bm{\xi},t) \,,
\label{Eq:GPE-final}
\end{multline}
where $\mu_j$ is the chemical potential associated with the species $j$ at time $t=0$ and where $\bm{\Lambda}^{\!-2}$ is a notation for the diagonal matrix $[\bm{\Lambda}^{-1}\bm{\Lambda}^{-1}]$. Note that for the reference species $j=j^*$, this equation simplifies to
\begin{multline}
i\hbar\,\partial_t\Psi_{j^*}^S(\bm{\xi},t) = \bigg[
- \dfrac{\hbar^2}{2m_{j^*}}
\bm{\nabla}_{\!\bm{\xi}}^T\bm{\Lambda}^{\!-2\,}\bm{\nabla}_{\!\bm{\xi}}
+ \overline{W}_{\!j^*}(\bm{\Lambda}\,\bm{\xi},t)\\
+ \frac{\frac{m_{j^*}}{2}\,\bm{\xi}^T\bm{\Omega}_{j^*}^{\prime\,2}(0)\,\bm{\xi}
+ \sum_{j'}N_{j'}\,g_{j^*j'}|\,\Psi_{j'}^S(\bm{\xi},t)|^2 - \mu_{j^*}}{\det[\bm{\Lambda}(t)]} \bigg]\,\Psi_{j^*}^S(\bm{\xi},t) \,.
\label{Eq:GPE-final-simple}
\end{multline}

These series (\ref{Eq:GPE-final}) and (\ref{Eq:GPE-final-simple}) of coupled differential equations, which constitute the main result of this paper, are solved numerically using the second-order split-operator technique \cite{Feit1982}. This technique is first used in imaginary time \cite{Lehtovaara2007, Bao2004} to compute the ground state of the binary mixture, which is taken as the initial state of the system at time $t=0$. It is then used in real time to compute the temporal dynamics of the system \cite{CorgierNJP2018, CorgierNJP2020}. Although in a multi-species mixture the individual species $j$ are typically trapped in potentials with different trap frequencies leading to unequal expansion dynamics, the scaling introduced in Eqs.\,(\ref{Eq:GPE-final}) and (\ref{Eq:GPE-final-simple}) still absorbs most of the dynamics such that the numerical solution of the time evolution can be obtained much faster compared with a static grid. Moreover, in the special case of equal trap frequencies for both species, which could be realized with dedicated optical traps \cite{MeisterQST2023}, Eqs. (\ref{Eq:GPE-final}) and (\ref{Eq:GPE-final-simple}) further simplify.

\section{Applications}
\label{sec:applications}

In sections \ref{sec:Transport} and \ref{sec:Expansion} we discuss two typical examples of the dynamics of a binary mixture of K-41 and Rb-87 that strongly benefit from applying our scaling techniques for an efficient numerical simulation. Furthermore, in section \ref{sec:MAIUS} we present a direct comparison of this theoretical approach with experimental measurements recently carried out on ground with the payload of the sounding rocket MAIUS-2.

\subsection{Transporting a Binary Mixture in Microgravity}
\label{sec:Transport}

The first example consists of a transport of the mixture confined on an atom chip by shifting the trap minimum over a distance of about 20\,\textmu{}m in 10 ms, followed by a holding period of 20\,ms in the final trap. Throughout the transport duration, we assume that the trap remains almost cylindrically symmetric, and the size of the condensate varies only slightly. Such transport protocols are mandatory for preparing the mixture as a source for subsequent atom interferometry measurements for conducting a test of the universality of free fall \cite{BattelierExpAstro2021}, where transports up to millimeter distances are needed \cite{CorgierNJP2018,GebbeNatComm2021}.

\subsubsection{Sequence details:}

We consider that the atoms are trapped by the magnetic field produced by a Z-shaped atom chip configuration \cite{Hansel2001, Folman2002, Nirrengarten2006, Reichel2011, CorgierNJP2018} in the presence of a time-dependent homogeneous magnetic field generated by magnetic coils through which flows a tunable current. The transport dynamics considered in this example is induced by a linear variation of the coil current during 10\,ms. Since the relative variation of this current remains small, the trajectory followed by the center of the trap during these 10\,ms is also linear and it is uniform, and the evolution of the trapping frequencies over time is also linear. The dynamics is assumed to take place in microgravity, and the position of the center of the trap is therefore the same for potassium and rubidium. At time $t=0$ its initial position is $314.97$\,\textmu{}m above the atom chip. The transport consists of a translation in the $z$ direction, perpendicular to the chip, bringing the center of the trap to the distance $z = 333.56$\,\textmu{}m from the chip. The total length of the transport is thus $18.59$\,\textmu{}m, to be compared with the initial width (FWHM) of the atomic density distribution along $z$ of about $2$\,\textmu{}m. In the following, we will associate index 1 with rubidium and index 2 with potassium. For rubidium, the trapping frequencies vary from
\begin{subequations}
\label{Eq:freq0}
\begin{align}
\omega_{1,X_1}(0) & = 2\pi \times ~24.8\,\mathrm{Hz} \\
\omega_{1,Y_1}(0) & = 2\pi \times 378.3\,\mathrm{Hz} \\
\omega_{1,Z_1}(0) & = 2\pi \times 384.0\,\mathrm{Hz}
\end{align}
\end{subequations}
to
\begin{subequations}
\label{Eq:freq1}
\begin{align}
\omega_{1,X_1}(t_f) & = 2\pi \times ~24.9\,\mathrm{Hz} \\
\omega_{1,Y_1}(t_f) & = 2\pi \times 340.9\,\mathrm{Hz} \\
\omega_{1,Z_1}(t_f) & = 2\pi \times 346.4\,\mathrm{Hz}
\end{align}
\end{subequations}
The initial and final trapping frequencies $\omega_{2,\Sigma_2}$ for potassium are given by the relation
\begin{equation}
\label{Eq:freq2_freq1_ratio}
\omega_{2,\Sigma_2}(t) = \left(\frac{m_1}{m_2}\right)^{\!\!\frac{1}{2}} \omega_{1,\Sigma_1}(t)
\end{equation}
valid for magnetic trapping with $\Sigma = X$, $Y$ or $Z$.

\begin{figure}[!t]
\centering
\includegraphics[width=0.85\textwidth]{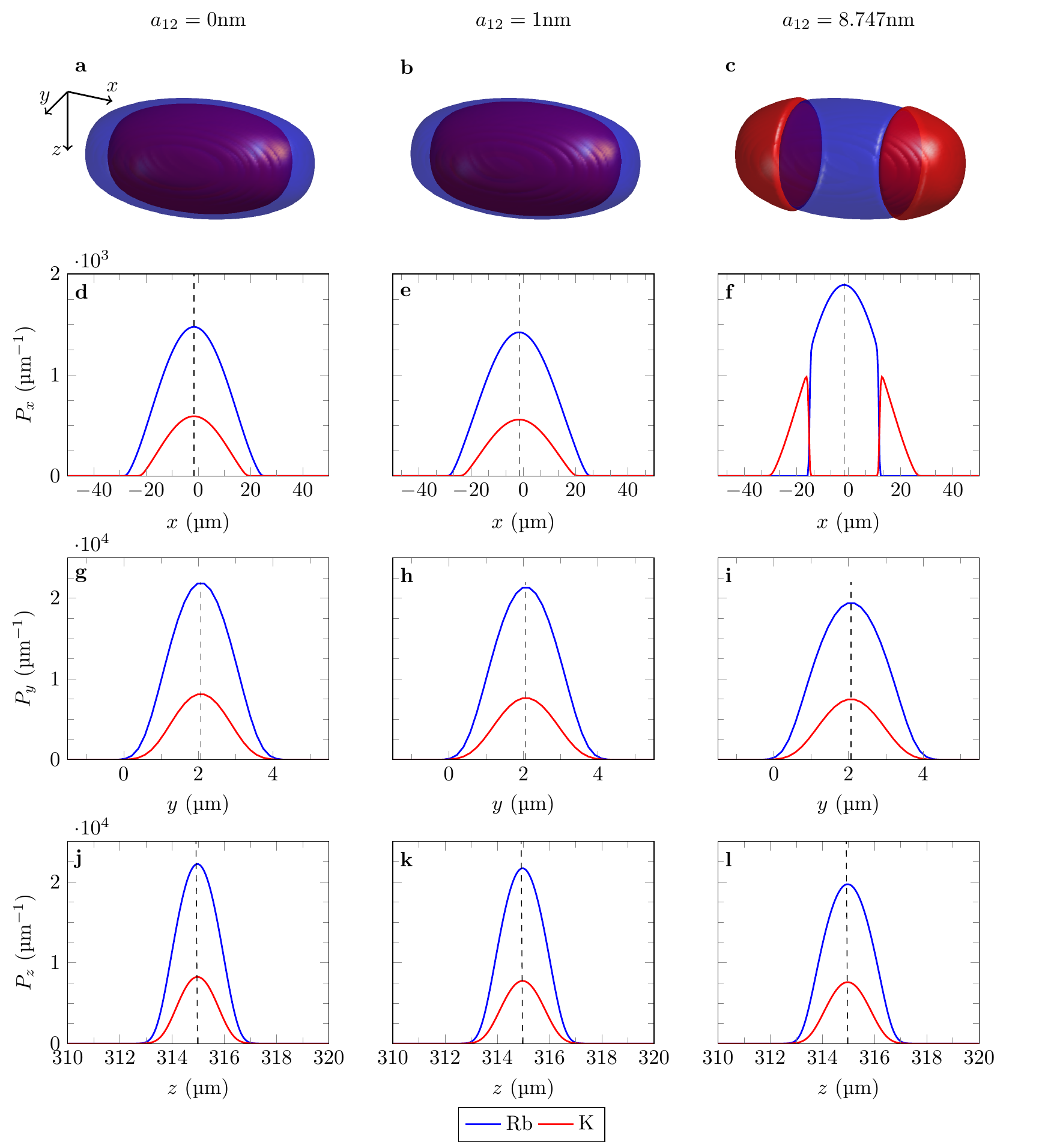}
\caption{Calculated ground state of a dual Rb-K condensate in microgravity in two miscible phases (left and central columns) and one immiscible phase (right column), in a cigar-shaped trap. The 3D representations are shown in the first row (panels \textbf{a}, \textbf{b} and \textbf{c}) and the integrated longitudinal and transverse density profiles $P_x$, $P_y$ and $P_z$ for Rb and K are shown in the next three rows (along $x$: panels \textbf{d}, \textbf{e}, \textbf{f} ; along $y$: panels \textbf{g}, \textbf{h}, \textbf{i} ; along $z$: panels \textbf{j}, \textbf{k}, \textbf{l}). The density profiles of rubidium and potassium are shown in blue and red, respectively. The intraspecies interaction parameters are $a_{11} = 5.237$\,nm and $a_{22} = 3.204$\,nm. The interspecies scattering length is $a_{12} = 0$ in the left column, $a_{12} = 1$\,nm in the central column and $a_{12} = 8.747$\,nm in the right column. The trap frequencies are given in Eqs.\;(\ref{Eq:freq0}) and (\ref{Eq:freq2_freq1_ratio}). The number of rubidium and potassium atoms are 43,900 and 14,400, respectively. The center of the trap is marked in each subplot by a black vertical dotted line.}
\label{fig:TransportGround}
\end{figure}

We consider a binary mixture of 43,900 rubidium atoms with 14,400 potassium atoms similar to what can be achieved regularly with the MAIUS-2 experiment on ground \cite{PiestThesis2021}. To explore different miscibility regimes, we consider the case of 3 values of the interspecies scattering length $a_{12} = 0$\,nm, $1$\,nm or $8.747$\,nm. This variation of the scattering length can, in principle, be realized experimentally using the Feshbach resonances observed in K-41 and Rb-87 mixtures around 35\,G and 79\,G using a dipole trap \cite{ThalhammerPRL2008}. The last value $a_{12} = 8.747$\,nm corresponds to the natural scattering length between K-41 and Rb-87 in the absence of any Feshbach resonance.

\subsubsection{Ground state:}

Before the transport dynamics of this double-species condensate can be studied, it is necessary to determine the steady state of the binary mixture confined in the initial trap. The ground state of this quantum mixture depends non-trivially on the respective strengths of the inter-species and intra-species interactions, which condition the miscibility of the two quantum fluids \cite{Ao1998, Timmermans1998, TrippenbachJPhysB2000}. This dependence is illustrated in Fig.\;\ref{fig:TransportGround}, which shows the influence of the value of the interspecies scattering length $a_{12}$ on the spatial distribution of the ground state atomic density obtained by solving the coupled Gross-Pitaevskii equation (\ref{eq_TDCGPE}) in imaginary time  \cite{Lehtovaara2007, Bao2004}. The first two columns correspond to two miscible cases associated with $a_{12} = 0$ and $a_{12} = 1$\,nm, respectively. The third column corresponds to the immiscible case $a_{12} = 8.747$\,nm which fulfills the immiscibility condition $g_{12}^2 > g_{11} g_{22}$ \cite{pethick_smith_2008,FerlainoPRA2006}. The first row shows a 3D representation of the atomic density associated with Rb (blue) and K (red). The immiscible nature of the mixture in the $a_{12} = 8.727$\,nm case is clearly visible in this 3D representation, which shows a discriminating hamburger-like structure. In contrast to the separation observed in this case between K and Rb, the two miscible cases are characterized by a large spatial overlap of the two condensates. The last three rows in Fig.\;\ref{fig:TransportGround} show the average atomic densities
\begin{subequations}
\begin{eqnarray}
P_{x}(x,t) & = & \int_{-\infty}^{\infty}\int_{-\infty}^{\infty}N_j\,|\Psi_j(\mathbf{r},t)|^2\;dy\,dz\\
P_{y}(y,t) & = & \int_{-\infty}^{\infty}\int_{-\infty}^{\infty}N_j\,|\Psi_j(\mathbf{r},t)|^2\;dx\,dz\\
P_{z}(z,t) & = & \int_{-\infty}^{\infty}\int_{-\infty}^{\infty}N_j\,|\Psi_j(\mathbf{r},t)|^2\;dx\,dy
\end{eqnarray}
\end{subequations}
for Rb (blue) and K (red) along the three directions $x$, $y$ and $z$ at initial time $t=0$. These plots lead to the conclusion that the two miscible cases considered here are very similar. Hence, compared to the non-interacting case ($a_{12} = 0$, left column of Fig.\;\ref{fig:TransportGround}), the introduction of a weak repulsive interaction between Rb and K ($a_{12} = 1$\;nm, central column of Fig.\;\ref{fig:TransportGround}) has very little impact on the initial spatial distribution of the atomic densities. In comparison with these miscible cases, the non-miscible case shows spatial distributions along $y$ and $z$ (panels \textbf{i} and \textbf{l}, right column in Fig.\;\ref{fig:TransportGround}) that are relatively unaffected by the introduction of a strong repulsion between Rb and K atoms, with $a_{12} = 8.747$\;nm. The spatial discrimination is only observed in the direction of the weak axis of trapping, \textit{i.e.} in the $x$ direction (see panel \textbf{f} in Fig.\;\ref{fig:TransportGround}).

\subsubsection{Transport dynamics:}

\begin{figure}[!t]
\centering
\includegraphics[width=0.85\textwidth]{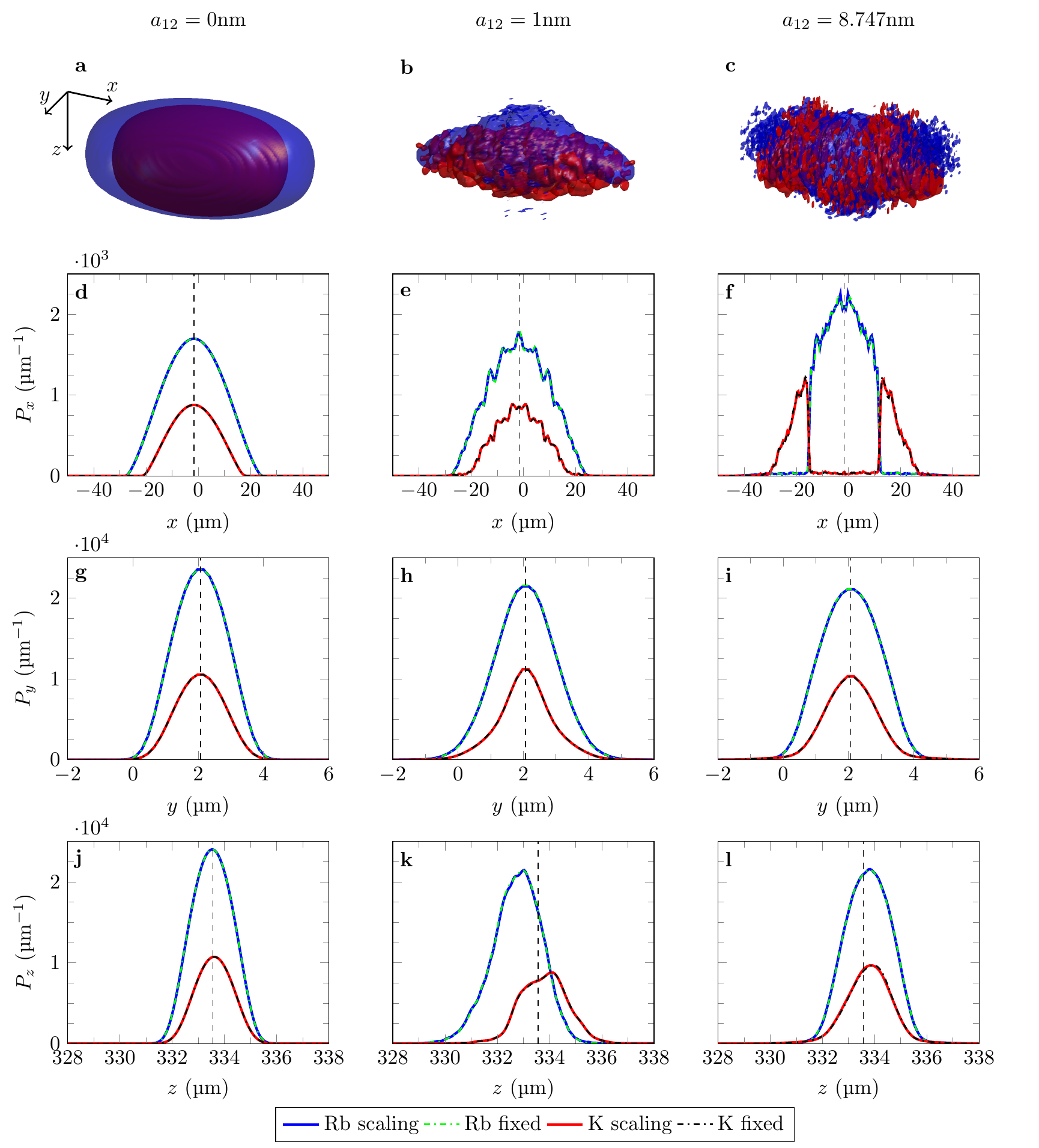}
\caption{Probability density of a dual Rb-K condensate in microgravity after 10\,ms of transport and 20\,ms of holding. Shown are the 3D representations (first row) and the integrated longitudinal and transverse density profiles $P_x$, $P_y$ and $P_z$ for Rb and K (next 3 rows). The interspecies scattering length is $a_{12} = 0$ in the left column, $a_{12} = 1$\,nm in the central column and $a_{12} = 8.747$\,nm in the right column. The number of Rb and K atoms are 43,900 and 14,400, respectively. The center of the trap is marked in each subplot by a black vertical dotted line. In panels \textbf{d} to \textbf{l}, the Rb probability densities calculated with the present grid-scaling approach and with a fixed grid are shown as solid blue lines and dashed green lines, respectively. Similarly, the K probability densities calculated with the present grid-scaling approach and with a fixed grid are shown as solid red lines and dashed black lines, respectively.}
\label{fig:TransportFinal}
\end{figure}

Fig.\;\ref{fig:TransportFinal} shows the calculated probability densities for Rb and K at the end of the transport and holding, at time $t=t_f=30$\,ms. The first row shows a 3D representation of the atomic densities. The next three rows show the averaged probability densities $P_x$, $P_y$ and $P_z$ calculated with the present grid-scaling method (blue and red solid lines for Rb and K, respectively) and with a fixed grid (green and black dashed lines for Rb and K, respectively). The probability densities calculated with these two different methods are perfectly superimposed, demonstrating the validity of the grid-scaling approach, whatever the chosen interaction regime, whether the mixture is miscible or not. It can be noted that if the results of these two approaches are identical, it is because these two methods are mathematically equivalent and therefore they can, in principle, differ only by the numerical errors induced by the limited precision of the calculations. A comparison of the panels \textbf{d} and \textbf{e} in Fig.\;\ref{fig:TransportFinal} also shows that the introduction of a weak interaction between Rb and K ($a_{12}=1$\,nm in panel \textbf{e}, 0 in panel\;\textbf{d}) induces significant perturbations in the spatial density profile in the $x$ direction corresponding to the weakest trapping axis, whereas the effect of these interactions was negligible in the ground state (see panels \textbf{d} and \textbf{e} in Fig.\;\ref{fig:TransportGround}). We can therefore conclude that the transport  acts here as a detector of these interspecies interactions, even if they are relatively weak. Comparing panels \textbf{j} and \textbf{k}, we also see that this weak interspecies interaction induces a shift in the average position of the two atomic species in the $z$-direction of transport, which is not the case in the absence of such interaction.

It can already be noted that in order to converge the calculation, it was necessary to use a larger number of grid points in the fixed-grid approach than in the grid-scaling approach. In fact, the fixed-grid approach uses $(N_x=256, N_y=64, N_z=576)$ grid points, while the grid-scaling approach uses $(N_x=256, N_y=64, N_z=192)$ grid points. The total number of $N_x \times N_y \times N_z$ grid points required is thus 3 times larger for the fixed-grid approach than for the grid-scaling approach. As shown in Table\;\ref{Tab:CPUt1}, this variation in the number of grid points obviously has a strong impact on the computational time. In fact, for the present calculation, the computation time, either for obtaining the fundamental state using the imaginary time approach \cite{Lehtovaara2007, Bao2004} or for calculating the dynamics, is on average 3 times longer with the fixed grid than with the grid-scaling approach. The ratio of 3 obtained here is due to the necessity of increasing the size of the grid in the $z$ direction, \textit{i.e.} in the direction in which the transport takes place. In this example it is limited to the value 3 because the transport achieved (with a displacement of about 18\,\textmu{}m) is not very large compared to the initial size of the condensate (about 2\,\textmu{}m FWHM in the $z$ direction, as shown in panels \textbf{j}, \textbf{k} and \textbf{l} of Fig.\,\ref{fig:TransportGround}). However, many experiments in the past have required the realization of condensate displacements over distances of the order of a millimeter \cite{Becker2018, Deppner2021, Gaaloul2022}. It can thus be estimated that the calculation of transport dynamics in such situations would require the use of 100 to 200 times more grid points in a fixed-grid calculation than in the grid-scaling approach, making this type of calculation extremely demanding in terms of memory resources as well as computational time, or even impossible with standard computing facilities.

\begin{table}[!t]
\caption{\label{Tab:CPUt1}Transport dynamics computation (cpu) time. The calculations were performed parallelizing 16 cores of an Intel Xeon Gold 6230 processor running at 2.1 GHz. The real calculation time is roughly the displayed values of the table divided by the number of cores.} 
\begin{indented}
\lineup
\item[]\begin{tabular}{ccccc}
\br
              & \centre{2}{Ground State Calculation} & \centre{2}{Ramp \& Holding Dynamics}\\
\ns
              & \crule{2}                            & \crule{2}                    \\
$a_{12}$ (nm) & ~~Fixed Grid~~ & ~~Scaled Grid~~ & ~~Fixed Grid~~ & ~~Scaled Grid~~ \\
\mr
0             & ~9\,h\;10\,min &  3\,h\;02\,min  &  9\,h\;25\,min & 3\,h\;03\,min \\
1             & ~9\,h\;39\,min &  3\,h\;00\,min  &  9\,h\;37\,min & 2\,h\;58\,min \\
8.747         & 24\,h\;19\,min &  8\,h\;03\,min  &  9\,h\;23\,min & 3\,h\;04\,min \\
\br
\end{tabular}
\end{indented}
\end{table}

\begin{figure}[!t]
\centering
\includegraphics[width=0.85\textwidth]{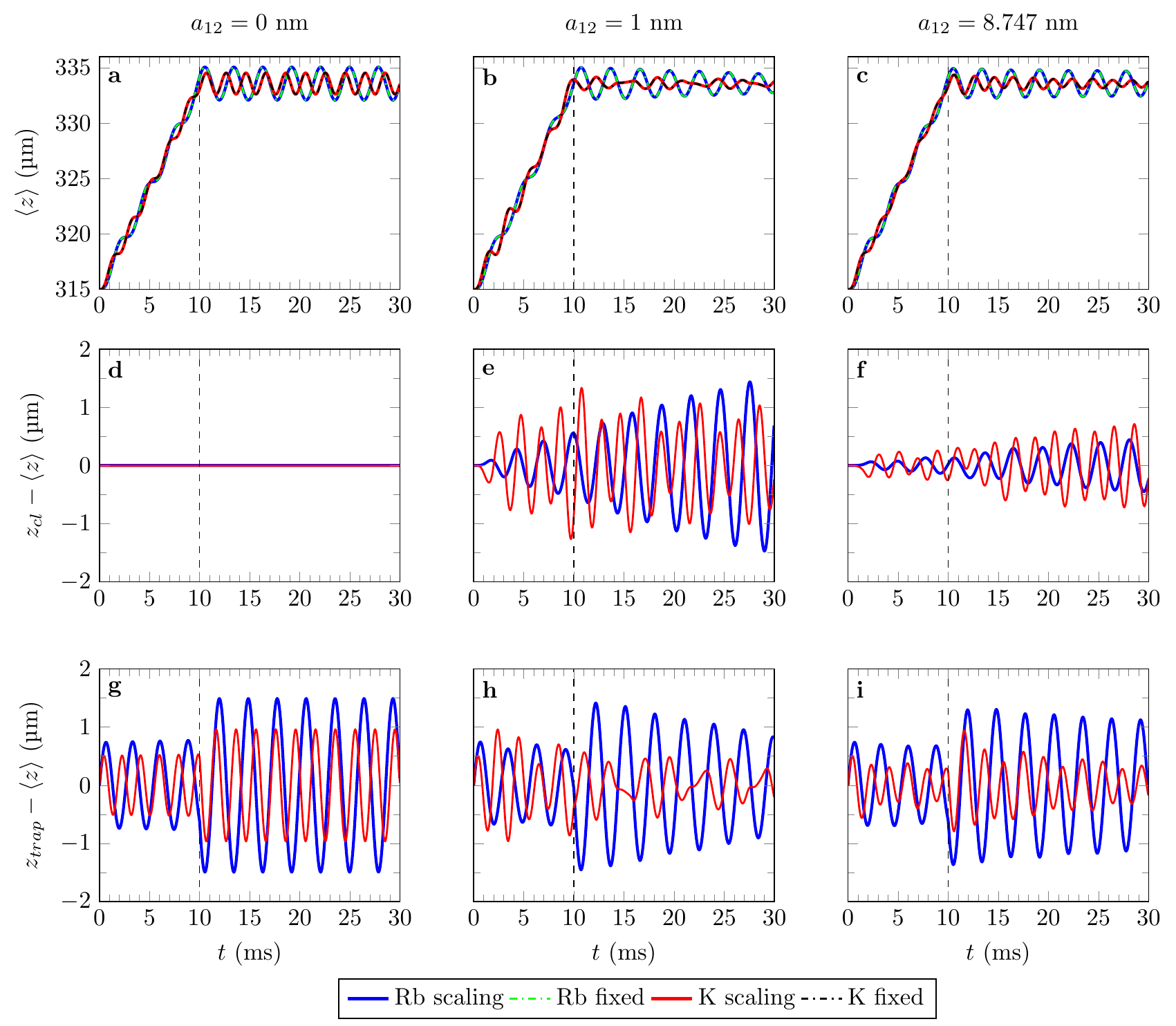}
\caption{Evolution of the atomic clouds average positions along $z$ during the transport and holding dynamics for the  3 values of interspecies scattering lengths considered here: $a_{12}=0$ left column, $a_{12}=1$\,nm central column, and $a_{12}=8.747$\,nm right column. The first row shows the average position $\langle z \rangle$ for Rb and K as a function of time. The second row shows the offset between this average position $\langle z \rangle$ and the trajectory $z_{cl}(t)$ expected if Newton's law applied independently for each species. The third row shows the offset between the average position $\langle z \rangle$ and the center of the trap. The color code associated with Rb and K is the same as in Fig.\;\ref{fig:TransportFinal}. The end of the transport and the beginning of the holding time is marked in each subplot by a black vertical dotted line.}
\label{fig:TransportPosEvolution}
\end{figure}

We will now present a more precise study of the displacement dynamics by calculating the average positions of the two condensates over time (quantities that we will consider as the ``trajectories'' followed by the two atomic clouds), and by calculating the evolution of the average ``size'' of the two condensates, defined as the standard deviations of the Rb and K atomic densities. The average trajectories followed by the two condensates defined as
\begin{equation}
\langle z \rangle = \iiint 
\Psi_j^*(\mathbf{r},t) \,z\, \Psi_j(\mathbf{r},t)\, d\mathbf{r}
\end{equation}
 are shown in the first row of Fig.\;\ref{fig:TransportPosEvolution} using the same color coding as in Fig.\;\ref{fig:TransportFinal}. We see in the panels \textbf{a}, \textbf{b} and \textbf{c} that the fixed-grid calculations and the grid-scaling approach give the same results regardless of the interaction regime considered. It can be seen in panels \textbf{g}, \textbf{h} and \textbf{i} of Fig.\;\ref{fig:TransportPosEvolution} that the condensates start to oscillate in their respective potential wells from the beginning of the transport. This is because the transport is too fast to be adiabatic. Furthermore, these oscillations, which occur at different frequencies for Rb and K, continue into the holding phase. When Rb and K do not interact, we see in panel \textbf{g} that the two condensates collide at regular time intervals. In the presence of interspecies interactions these collisions strongly perturb the trajectories followed by the two condensates. Consequently, even if the average positions of the two condensates obey the classical laws of motion when the interspecies interaction is suppressed (see panel \textbf{d}), this is no longer the case in the presence of an interaction (see panel \textbf{f}), even if this interaction is relatively weak (see panel \textbf{e}). Finally, panels \textbf{g}, \textbf{h} and \textbf{i} show that the remaining oscillations observed in the holding phase are characterized by multiple modes that differ as a function of the interspecies scattering length.

\begin{figure}[!t]
\centering
\includegraphics[width=0.85\textwidth]{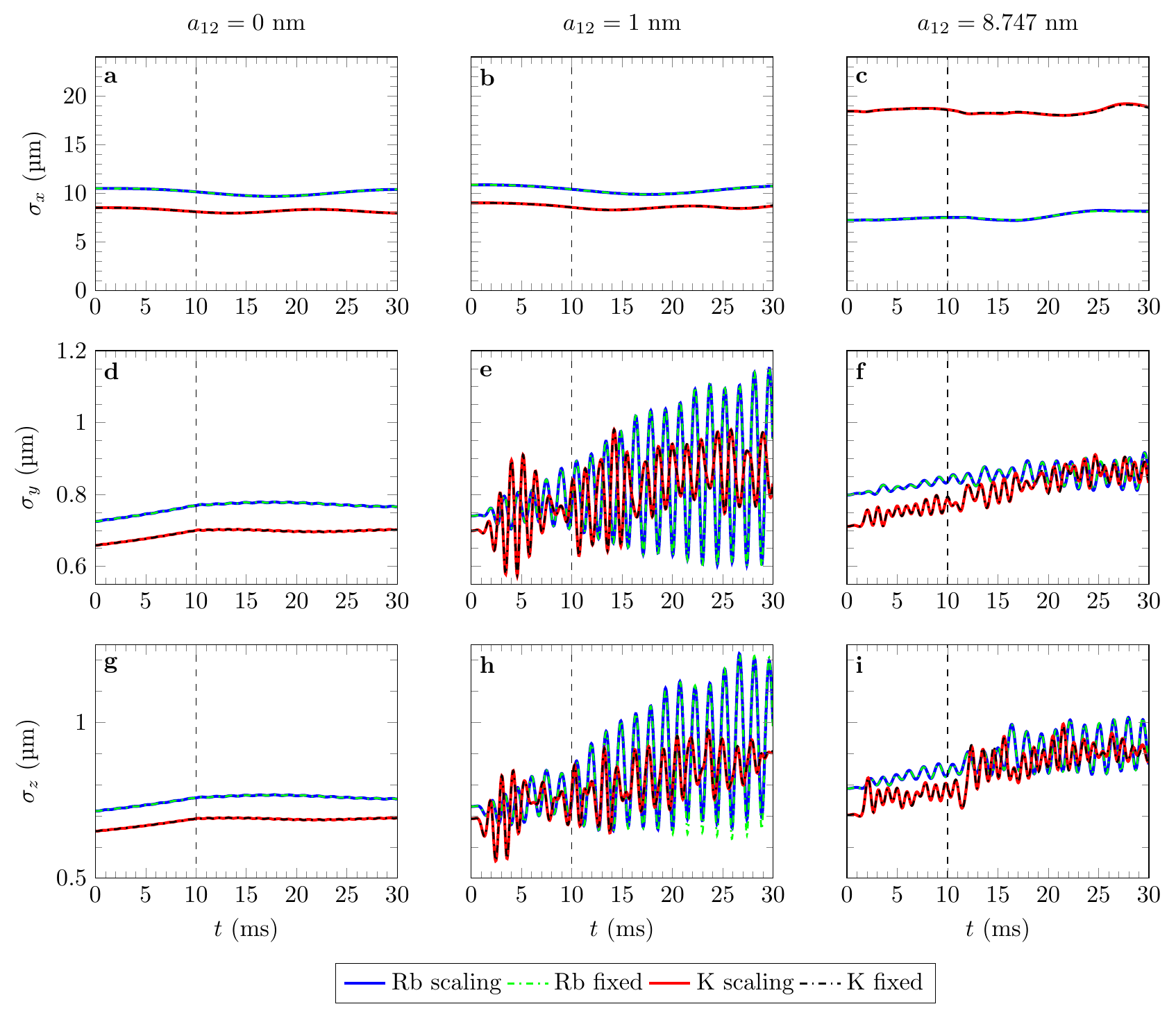}
\caption{Evolution of the size (atomic density standard deviation) of the Rb and K clouds along $x$, $y$ and $z$ during the transport and holding dynamics for the 3 values of interspecies scattering lengths considered here: $a_{12}=0$ left column, $a_{12}=1$\,nm central column, and $a_{12}=8.747$\,nm right column. The color code associated with Rb and K is the same as in Fig.\;\ref{fig:TransportFinal}. The end of the transport and the beginning of the holding time is marked in each subplot by a black vertical dotted line.}
\label{fig:TransportSizeDyn}
\end{figure}

The evolution of the average sizes of the two condensates, defined as the standard deviations of the Rb and K atomic densities along $x$, $y$ and $z$, are shown in Fig.\;\ref{fig:TransportSizeDyn} using the same color coding as in Fig.\;\ref{fig:TransportFinal}. We see in this figure that also for the evolution of the sizes, the fixed-grid calculations and the multi-species grid-scaling approach presented here give identical results, regardless of the interaction regime. During transport, the trapping frequencies for Rb and K in the $x$ direction remain nearly constant, while the trapping frequencies along the $y$ and $z$ axes decrease by slightly less than 10\,\%. This relatively small evolution of the trap frequencies during the transport leads to a smooth evolution of the size of the two atomic clouds when the interspecies interaction is absent ($a_{12}=0$, left column of Fig.\;\ref{fig:TransportSizeDyn}). In contrast, panels \textbf{e}, \textbf{f}, \textbf{h} and \textbf{i} in Fig.\;\ref{fig:TransportSizeDyn} show that the presence of a non-zero interspecies interaction ($a_{12}=1$\,nm in the middle column and $a_{12}=8.747$\,nm in the right column) leads to relatively strong collective excitations of the two condensates in the $y$ and $z$ directions, which continue into the holding phase. Since the change of the trap frequencies along the $x$ direction is close to zero, no such perturbation effect is observed in this particular direction (see panels \textbf{b} and \textbf{c} in Fig.\;\ref{fig:TransportSizeDyn}).

\subsection{Free Expansion of a Binary Mixture under Gravity}
\label{sec:Expansion}

The second exemplary application of this multi-species grid-scaling approach is a free expansion of a binary Rb-K mixture in the presence of gravity. The number of atoms considered is again 43,900 for Rb and 14,400 for K. We simulate the free expansion of the Rb-K mixture, starting at $t=0$ from the ground state of this binary mixture. Due to the gravitational sag, the centers of the trapping potentials associated with each species are shifted, mainly in the $z$ direction, which is the direction in which gravity acts. In addition, the eigenaxes of the traps associated with Rb and K are very slightly rotated. The initial trap uses the same electric current flowing through the magnetic coil as in the example presented earlier in section\;\ref{sec:Transport}, which discussed the dynamics of transport and holding in microgravity. In the case of Rb, the initial trap is positioned at $x=-1.62$\,\textmu{}m, $y=2.23$\,\textmu{}m, $z=332.43$\,\textmu{}m. The trapping frequencies are
\begin{subequations}
\label{Eq:freq2}
\begin{align}
\omega_{1,X_1}(0) & = 2\pi \times ~25.3\,\mathrm{Hz}\\
\omega_{1,Y_1}(0) & = 2\pi \times 345.1\,\mathrm{Hz}\\
\omega_{1,Z_1}(0) & = 2\pi \times 347.1\,\mathrm{Hz}
\end{align}
\end{subequations}
For K, the initial trap is positioned around $x=-1.76$\,\textmu{}m, $y=2.24$\,\textmu{}m, $z=331.35$\,\textmu{}m, and the trapping frequencies are
\begin{subequations}
\label{Eq:freq3}
\begin{align}
\omega_{2,X_2}(0) & = 2\pi \times ~36.5\,\mathrm{Hz}\\
\omega_{2,Y_2}(0) & = 2\pi \times 504.1\,\mathrm{Hz}\\
\omega_{2,Z_2}(0) & = 2\pi \times 509.8\,\mathrm{Hz}
\end{align}
\end{subequations}

\begin{figure}[!t]
\centering
\includegraphics[width=0.99\textwidth]{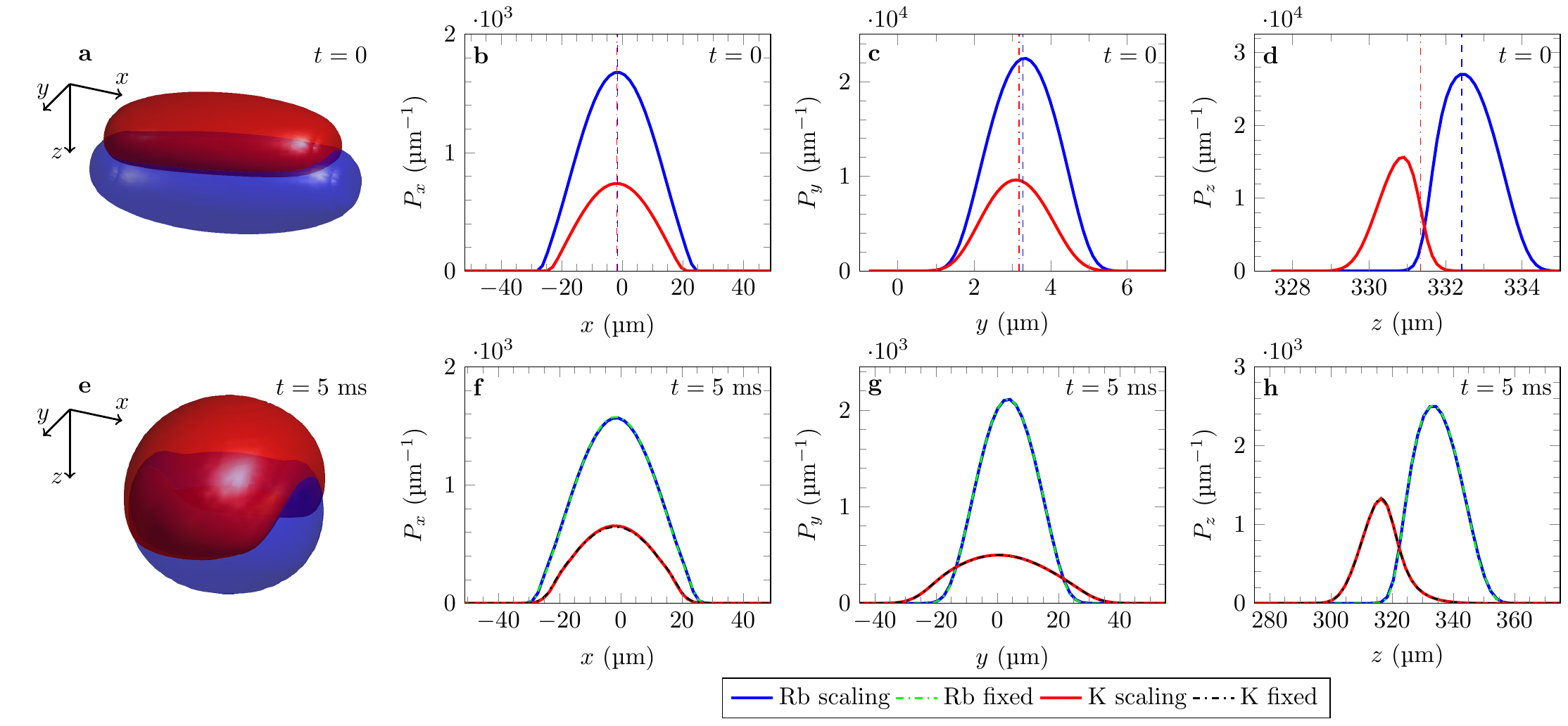}
\caption{Probability density of a dual Rb-K condensate in the presence of gravity. Shown are the 3D representations (first column) and the integrated longitudinal $P_x$ and transverse $P_y$ and $P_z$ density profiles for Rb and K (second column: along $x$, third column: along $y$, fourth column: along $z$). The first row shows the initial $(t=0)$ ground state, while the second row shows the same density after 5\,ms of free expansion. The intraspecies interaction parameters are $a_{11} = 5.237$\,nm and $a_{22} = 3.204$\,nm. The interspecies scattering length is $a_{12} = 8.747$\,nm. In the first row, the centers of the Rb and K traps are marked by blue and red vertical dotted lines, respectively. The trap frequencies at $t=0$ are given in the text. The number of rubidium and potassium atoms is 43,900 and 14,400, respectively. Gravity points in the positive $z$ direction. In the second row, the Rb probability densities calculated with the present grid-scaling approach and with a fixed grid are shown as solid blue lines and dashed green lines, respectively. Similarly, the K probability densities calculated with the present grid-scaling approach and with a fixed grid are shown as solid red lines and dashed black lines, respectively.}
\label{fig:FreeExp5ms}
\end{figure}

The first row of Fig.\,\ref{fig:FreeExp5ms} shows the spatial distribution of the dual-species condensate at time $t=0$, to be compared with the distribution shown in the last column of Fig.\,\ref{fig:TransportGround}, which shows the same data in a microgravity environment. From this comparison, we can already conclude that the presence of gravity significantly affects the initial structure of the condensate. The first notable change is that, in the presence of gravity, the symmetry of the hamburger-like structure of the condensate is broken. There are also significant areas where the two atomic clouds overlap. This was not the case in microgravity and this is due to the fact that in the presence of gravity the two traps are spatially offset from each other.

Since the size of the two-species condensate increases rapidly as the trap is released, we have limited the duration of the time-of-flight to 5\,ms only, so that a fixed grid calculation remains feasible. The second row of Fig.\,\ref{fig:FreeExp5ms} shows the spatial distribution of the dual-species condensate at time $t=5$\,ms, \textit{i.e.} at the end of this expansion.  It can be seen that during this time, the sizes of the Rb and K clouds typically grow by a factor of about 10 in both the $y$ and $z$ directions. On the contrary, in the weak axis direction $x$, the sizes of the clouds remain almost unchanged. Finally, the second row of Fig.\,\ref{fig:FreeExp5ms} compares the probability densities calculated at the end of the expansion with the present grid-scaling method (blue and red solid lines for Rb and K, respectively) with those obtained using a fixed grid (green and black dashed lines for Rb and K, respectively). The probability densities calculated with these two methods are in perfect agreement, thus confirming the validity of the grid-scaling approach in this example, where the atomic expansion dynamics occurs in the presence of gravity.

As with the transport and holding example discussed in the previous section \ref{sec:Transport}, describing the expansion with a fixed grid required a larger number of grid points than with the grid-scaling approach in order to achieve convergence. In fact, in this example, the fixed-grid approach uses ($N_x = 64$, $N_y = 256$, $N_z = 256$) grid points, while the grid-scaling approach uses ($N_x = 64$, $N_y = 64$, $N_z = 64$) grid points. The total number of $N_x \times N_y \times N_z$ grid points required is therefore 16 times greater for the fixed-grid calculation (4,194,304) than for the grid-scaling approach (262,144). As shown in Table\;\ref{Tab:CPUt2}, this variation in the number of grid points has a strong impact on the computation time, dramatically favouring the grid-scaling approach in terms of both CPU time and, of course, memory consumption. One can note that the increase in the number of grid points affects the $y$ and $z$ directions along which the condensate expansion is most significant in the first 5\,ms. The computation time, both for obtaining the ground state and for computing the dynamics, is on average 18 times larger with the fixed grid than with the grid-scaling approach, which is consistent with the ratio of grid sizes. Of course, this factor of 18 depends on the expansion time, since the size of the condensate increases linearly with time after the initial acceleration phase. As shown in Table\;\ref{Tab:CPUt2}, for an expansion time of 8\,ms, the computation time in a fixed grid is on average 68 times larger than in the grid-scaling approach. In fact, this computation requires $64 \times 512 \times 512$ grid points, \textit{i.e.} 64 times more than with the grid-scaling approach. In practice, many free expansion experiments are performed over durations of several tens of milliseconds \cite{Becker2018, Deppner2021, Gaaloul2022}. A simple extrapolation of the results obtained here gives a gain in computational time of the order of 600 for a free expansion of 25\,ms and of 10,000 for a time of flight of 100\,ms. Such calculations quickly become cumbersome in the standard fixed-grid approach, which confirms the importance of developing the grid-scaling approach proposed here for an efficient treatment of the expansion dynamics of multispecies quantum mixtures. Reaching the regime of few seconds of free expansion is also within reach since the scaled grid calculation time scales linearly with the expansion time and would amount to less than one hour of real computation time for 1s (last row of table 2).

\begin{table}[!t]
\caption{\label{Tab:CPUt2}Computation (cpu) time for the calculation of a dual-species condensate free expansion dynamics in gravity. The calculations were performed parallelizing 16 cores of an Intel Xeon Gold 6230 processor running at 2.1 GHz. The real calculation time is roughly the displayed values of the table divided by the number of cores.}
\begin{indented}
\lineup
\item[]\begin{tabular}{ccccc}
\br
         & \centre{2}{Ground State}          & \centre{2}{Expansion Dynamics}      \\
\ns
         & \crule{2}                         & \crule{2}                           \\
TOF (ms) & ~~Fixed Grid~~  & ~~Scaled Grid~~ & ~~Fixed Grid~~  & ~~Scaled Grid~~ \\
\mr
5        & 11\,h\;06\,min  & 31\,min\;00\,s  & 55\,min\;00\,s  & ~\,3\,min 40\,s  \\
8        & 39\,h\;07\,min  & 31\,min\;00\,s  & 6\,h\;12\,min   & ~\,6\,min 06\,s  \\
25       & 16\,days$^{*}$  & 31\,min\;00\,s  & 7\,days$^{*}$   & 18\,min 32\,s    \\
100      & 266\,days$^{*}$ & 31\,min\;00\,s  & 432\,days$^{*}$ & 78\,min 31\,s    \\
1000      & N/A & 31\,min\;00\,s  & N/A & 13\,h$^{*}$    \\
\br
\end{tabular}
\item[] $^{*}$ Estimation based on the number of grid points required.
\end{indented}
\end{table}

\subsection{Comparison with Experiment: Free Expansion of a Binary Mixture under Gravity}
\label{sec:MAIUS}

The successful launch of the MAIUS-1 mission led to the first demonstration of Bose-Einstein condensation in space \cite{Becker2018} and to the realization of the first interference experiments on board a sounding rocket \cite{Lachmann2021}. The MAIUS-2 and MAIUS-3 missions aim to study the dynamics of Rb-87 and K-41 mixtures in zero gravity and to prepare a quantum test of the universality of free fall in space. These missions have led to the development of a new atom chip device for trapping, condensing and manipulating Rb-87 and K-41 atoms together \cite{PiestThesis2021}. Using this setup, quantum degenerate mixtures with variable ratios of Rb to K atom numbers could be prepared, and this has led recently to the realization of several free expansion experiments of these binary mixtures on ground \cite{PiestThesis2021}. Here, we present a small subset of these results to verify the applicability of our computational method by comparing its predictions with experimental measurements.

\begin{figure}[!t]
\centering
\includegraphics[width=0.65\textwidth]{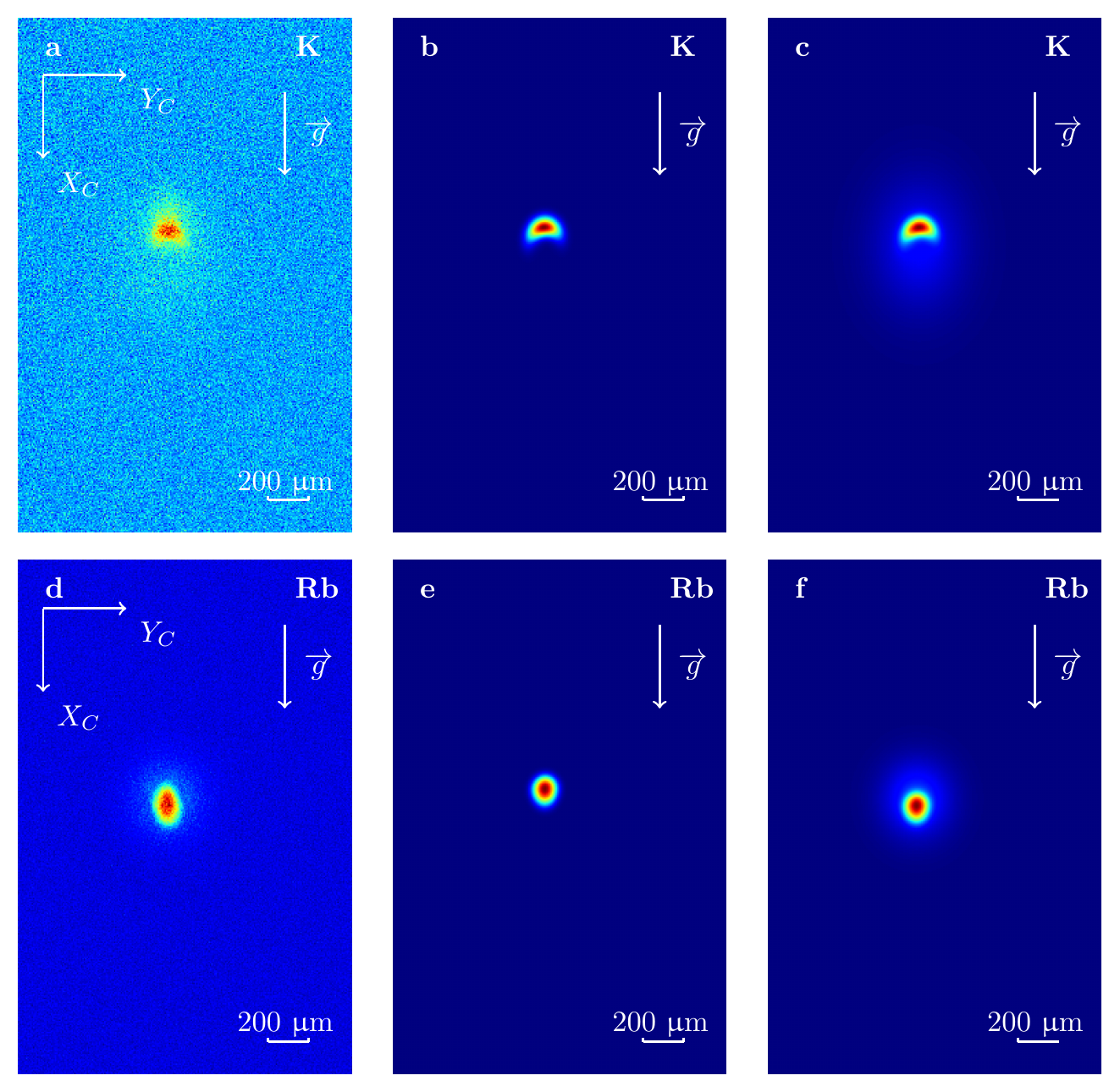}
\caption{Atomic densities of K (first row) and Rb (second row) after a free expansion of 25\,ms, starting from the initial trap described in section \ref{sec:Expansion}. First column: False-color absorption images measured by the MAIUS-2 apparatus in a ground-based experiment \cite{PiestThesis2021}. The direction of gravity, indicated by white arrows, is vertical, along the camera axis $X_C$, and the plane of the atom chip is perpendicular to gravity. The Rb and K images are normalized for better visibility. The fitted atom numbers are 43,900 for Rb and 14,400 for K \cite{PiestThesis2021}. Second column: Condensate probability densities calculated with the present grid-scaling approach in the plane $(X_C,Y_C)$ of the camera, after 25\,ms of free expansion. Third column: Calculated total probability densities, including thermal atoms.}
\label{fig:KRbCameraView}
\end{figure}

The first column of Fig.\;\ref{fig:KRbCameraView} shows measured absorption images of the K (first row, panel \textbf{a}) and Rb (second row, panel\;\textbf{d}) clouds after a free expansion of 25\,ms. The bright red regions correspond to density maxima and the dark blue regions to low atomic densities. The numbers of Rb and K atoms, calibrated by experimental measurements, are 43,900 and 14,400, respectively. Comparing panels \textbf{a} and \textbf{d} we can see that the experimental image of K is characterized by a background noise that is more important than for Rb because there are about 3 times less atoms of K than of Rb. The intensity of the peak is therefore lower for K than for Rb, and the signal-to-noise ratio is thus lower. The vertical direction $X_C$ of the camera corresponds to the direction $z$ of gravity. The horizontal axis $Y_C$ of the camera is in the $(x,y)$ plane, and makes an angle of 46\;degrees with the $x$-axis of Fig.\;\ref{fig:FreeExp5ms}. The initial Rb and K trapping frequencies are given in Eqs.\;(\ref{Eq:freq2}) and (\ref{Eq:freq3}) and the initial state of the condensed binary mixture has already been shown in the first row of Fig.\;\ref{fig:FreeExp5ms}. The second column of Fig.\;\ref{fig:KRbCameraView} shows the condensate atomic densities calculated after 25\,ms of free expansion by numerical solution of the coupled Gross-Pitaevskii Eqs.\;(\ref{eq_TDCGPE}) in the present grid-scaling approach, with K in the first row (panel \textbf{b}) and Rb in the second row (panel \textbf{e}). The grid used in the numerical calculation has been translated so that the position of the maximum K density is the same for the experimental and simulated data. A comparison of the Rb panels \textbf{d} and \textbf{e} then shows a slight shift between the measured position for the Rb cloud and its calculated position. This shift is about 81.6\,\textmu{}m in the $X_C$ direction of gravity and about 16.3\,\textmu{}m in the transverse $Y_C$ direction. Compared to the distance of 3,066\,\textmu{}m covered by the atoms during the 25 ms of free fall, this global shift of 83.2\,\textmu{}m between experiment and theory remains relatively limited, since it represents only 2.7\% of the total displacement. This small shift may be due to an initial oscillation of the atoms before the expansion stage in the experiment, or to an additional kick experienced by the atoms during the trap suppression, two effects that are not considered in the simulation. Nevertheless, it can be concluded that the comparison of the experimental measurements with the numerical simulation shows at this stage a good \textit{qualitative} agreement between theory and experiment in the region of interest captured by the CCD camera. It should also be noted that an efficient simulation of the 3D dynamics of the mixture was only possible by considering the scalings for the center of mass and the size expansion presented in sections \ref{sec:theory:com_scaling} and \ref{sec:theory:size_scaling}.

\begin{figure}[!t]
\centering
\includegraphics[width=0.75\textwidth]{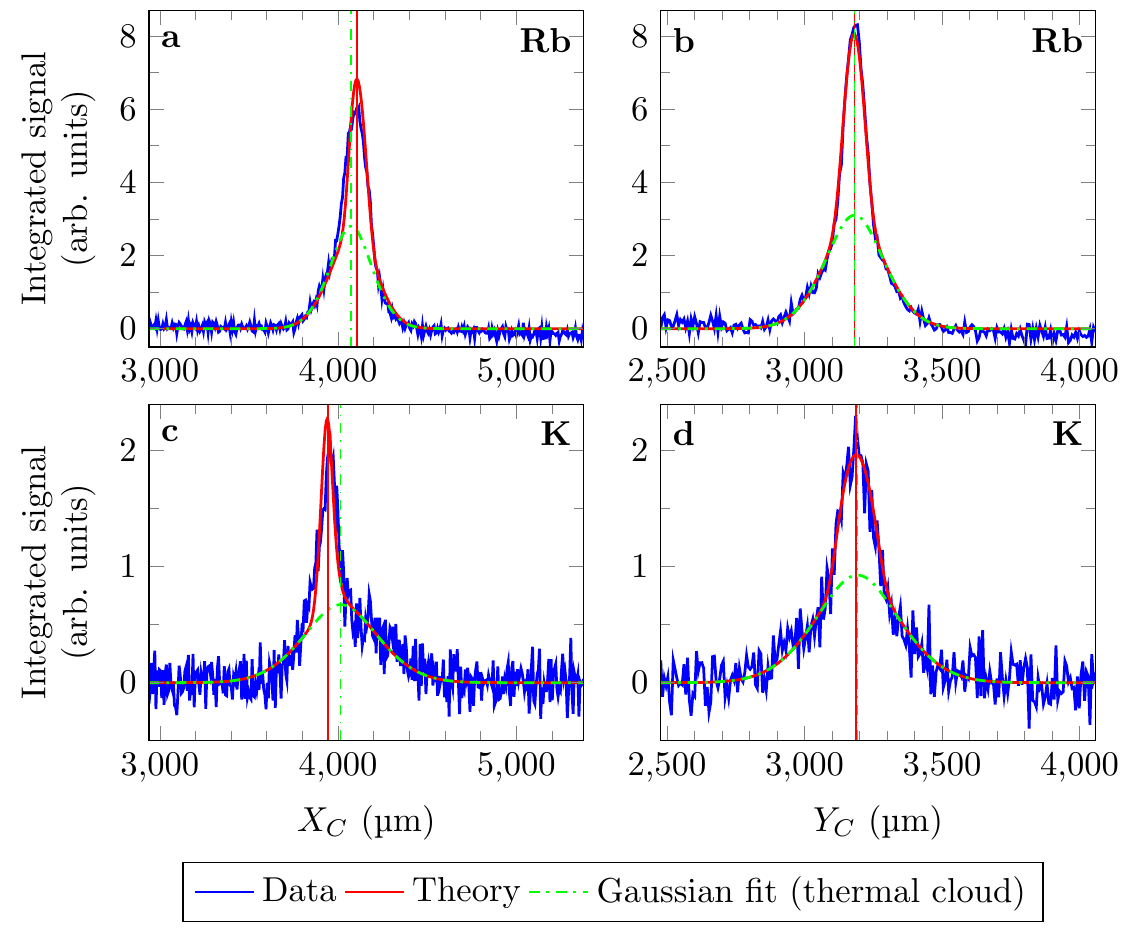}
\caption{Integrated atomic densities of Rb (first row) and K (second row) in arbitrary units. In the first column the integration of the 2D images shown in Fig.\;\ref{fig:KRbCameraView} is performed along $Y_C$ and in the second column the integration is performed along $X_C$. The solid blue line is the experimental measurement and the green dashed line is the Gaussian fit describing the thermal component of the atomic cloud. The solid red line is the adjusted numerical simulation, which includes both the computed condensed and the fitted thermal components. The vertical lines correspond to the average position of the condensed part of the atomic cloud for the solid red line, and to the centre of the thermal part for the dash-dotted green line. The offsets between the positions of the condensed and thermal parts in the directions $(X_C,Y_C)$ are $(+33.54$ \textmu m, $-0.41$ \textmu m) for Rb, and $(-40.93$ \textmu m, $-5.25$ \textmu m) for K.}
    \label{fig:Data25msTOF}
\end{figure}

A more quantitative study was then performed to refine this comparison. The size of the image shown in Fig.\;\ref{fig:KRbCameraView} corresponds to the region of interest taken for data analysis, and the intensity information given by the pixels of the camera was then integrated in each direction to obtain the integrated 1D signals shown as blue solid lines in Fig.\;\ref{fig:Data25msTOF}. These integrated experimental data are characterized by bimodal structures, with quasi-Gaussian pedestals corresponding to the presence of a thermal cloud. These pedestals observed for both Rb and K were fitted by 2-dimensional Gaussian functions using the 2D camera-recorded data shown in Fig.\;\ref{fig:KRbCameraView} \textbf{a} and \textbf{d}, and their integrals are shown as green dash-dotted lines in each subplot of Fig.\;\ref{fig:Data25msTOF}. The same integration is also performed on the simulated condensed atomic densities, followed by a numerical convolution by a Gaussian function with standard deviation (RMS width) $\sigma=15$\,\textmu{}m to mimic the effect of camera resolution. The simulated Rb peak was shifted by 81.6\,\textmu{}m in the $X_C$ direction and by 16.3\,\textmu{}m in the $Y_C$ direction, in agreement with the observations made previously in Fig.\;\ref{fig:KRbCameraView}. This shift was introduced to account for the initial velocity difference between Rb and K, which is not included in the simulation. Using this procedure, the simulated data describing the total density associated with the thermal and condensed atoms are finally plotted as solid red lines in Fig.\;\ref{fig:Data25msTOF}. The comparison between the experimental measurement (solid blue line) and the result of the numerical simulation using the grid-scaling approach presented here (solid red line) shows a very good agreement. The result of this numerical model taking into account simulated condensed and fitted thermal atoms is also shown in the right column of Fig.\;\ref{fig:KRbCameraView}, which also compares very favorably with the image captured by the camera (see left column of Fig.\;\ref{fig:KRbCameraView}). As a result, we conclude that our numerical approach enables efficient and accurate simulation of the dynamics of BEC mixtures in a wide range of realistic situations. It is worth noting that the centers of the condensed and thermal fraction distributions do not coincide, as is commonly expected, especially for the lighter K species (see Fig.\;\ref{fig:Data25msTOF}). This non-obvious effect is due to the repulsion between the dense, interacting, degenerate parts of the clouds, which causes a shift of the centre of each BEC with respect to its thermal counterpart. This can be seen as a signature of the bimodal distributions of interacting quantum mixtures.

\section{Conclusion}
\label{sec:conclusion}

This article presents an efficient method for describing the dynamics of quantum interacting mixtures. It is based on the translation and rescaling of the computational grid during the simulation of the coupled multi-species Gross-Pitaevskii equations. Perfect agreement with previous methods is shown in regimes where they could be computed. In addition, experimental validation was performed for time scales that would have been very challenging with previously used static-resolution grids. The validity of the developed approach allows its implementation in the context of microgravity and space experiments, where transports over long distances are realized at very low frequencies (a few Hz) and for long free expansion times of seconds, necessary in metrology applications such as fundamental physics tests \cite{AguileraClassQuGrav2014,AhlersArXiv2022stequest} or in the Earth observation context with space quantum gravimeters \cite{LevequeArXiv2022carioqa}. In these latter cases, our method would  take few hours whereas fixed-grid ones are not possible to implement at reasonable time scales.

\section{Acknowledgements}
The authors thank Gabriel Müller for his atom chip simulations and input of the trap configurations used in this paper. This work was funded by the Deutsche Forschungsgemeinschaft (German Research Foundation) under Germany’s Excellence Strategy (EXC-2123 QuantumFrontiers Grants No. 390837967) and through CRC 1227 (DQ-mat) within Projects No. A05, and the German Space Agency (DLR) with funds provided by the German Federal Ministry of Economic Affairs and Energy (German Federal Ministry of Education and Research (BMBF)) due to an enactment of the German Bundestag under Grants No. 50WM2250A (QUANTUS plus), No. 50WP1700 (BECCAL), No. 50WM2245A (CAL-II), No. 50WM2263A (CARIOQA-GE), No. 50WM2253A (AI-Quadrat), No. 50NA2106 (QGYRO+) and No. 50WP1431-1435 (QUANTUS-IV-MAIUS).

\section*{References}
\bibliography{biblio}

\end{document}